# PARADOXES OF THERMODYNAMICS AND STATISTICAL PHYSICS

DRAGOLJUB CUCIĆ

*Regional centre for talents Mihajlo Pupin, Pančevo, Serbia, rctpupin@panet.rs*

ABSTRACT

The paradoxes of thermodynamics and statistical physics are unavoidable in the study of physical paradoxes because of their importance at the time they came to be as well as the frequency of their appearance in historical studies of physics. In this work paradoxes are presented together with the historical studies of their creation, their solutions are given and they are analysed according to a number of characteristics: is it an exparadox, is it of theoretic or experimental nature, is it a thought experiment and the type of paradox it belongs to.

## Introduction

The subject of this work are paradoxes of thermodynamics and statistical physics. In the professional literature one can find a lot of texts dealing with the paradoxes in this branch of physics, but they are all solved and processed individually. In scientific works of this kind the solutions of each of the paradoxes are presented more thoroughly than here.

There is no text, or at least I haven't had the opportunity to find one, that synthesizes all more familiar paradoxes of thermodynamics and statistical physics according to common characteristics. This is the very aim of this work; to clearly number the paradoxes of thermodynamics and statistical physics that are considered by the author as worthy of attention; to show the historical conditions prior to paradox formulation; to give the current paradox solutions, if any exist, without in-depth explanations, since to solve certain paradoxes a very specific professional knowledge is needed and finally to gain a general insight into the paradoxes according to the following characteristics: is there a solution to the paradox, is it of theoretic or experimental nature, is the paradox a thought experiment and the paradox type.

This kind of analysis will provide a quantitative understanding and the opportunity to draw a conclusion concerning a possible special nature of paradoxes of thermodynamics and statistical physics. The results will be given, for better clarity, in graphical and tabular form.

This kind of analysis provides an approach to the phenomenon of paradox from the perspective of this branch of physics in order to get as complete picture of them as possible. A clear formulation of the causes of paradox in physics provides an easier path to their solutions if the paradoxes are equivalent in certain characteristics.

## Paradoxes of thermodynamics and statistical physics

Thermodynamics is a branch of physics that, like other branches of the science, has its origins in Ancient Greek philosophy. The breakthroughs in the study of heat phenomena date back to the middle ages with the attempts of scale standardisation and physical realisation of measuring device. The invention of the thermometer in the 17$^{th}$ century enables the heat to be studied quantitatively as well as qualitatively and thermodynamics begins clearly to take shape.

The first statistical data appeared in the 17$^{th}$ century but it is not until the 19$^{th}$ century that statistics and probability become inseparable parts of physics.
Thermodynamics and the statistical approach to thermodynamics (statistical physics) became, with the works of German scientist Rudolf Clausius (Rudolf Julius Emanuel Clausius, 1822–1888), Scotsman James Maxwell (James Clark Maxwell, 1831–1879), Austrian physicist Ludwig Boltzmann (Ludwig Eduard Boltzmann, 1844–1906) and Willard Gibbs (Willard Josiah Gibbs, 1839–1903), an objective physical entity – a view of the world.

The greatest number of statistical physics paradoxes originate from the 19$^{th}$ century, at the time of the formulation of statistical physics. The following paradoxes will be presented and analysed here:

1. Boltzmann's paradox.,
2. Loschmidt's reversibility paradox,
3. Zermelo's recurrent paradox,
4. Mpemba effect,
5. Maxwell's demon paradox,
6. Heat death paradox (Clausius paradox),
7. Gibbs' paradox (paradox of mixing).

*Gibbs' paradox* and *paradox of mixing* are often cited separately in literature. For example, in Wikipedia[1] it can be found that *Gibbs' paradox* is the question: "*In an ideal gas, is entropy an extensive variable?*", and paradox of mixing is:" *System entropy change before and after mixing.*" There is no essential difference between these sentences. This work will assume that these are just two names of the same paradox.

### 1. Boltzmann's paradox

*How is it that reversible microscopic movement of atoms leads to irreversible macroscopic phenomena?*

Maybe the simplest way to present the difference between micro and macro systems would be in "film" terms, since modern generations communicate and think mostly in this "media" language. A microsystem would be a "filmed" system in which we wouldn't be able to ascertain the direction the film was going in, while a macrosystem would easily show "inconsistencies" that would enable us to deduce the direction of the film. The judgment on the accuracy of the sequence of events is made based on the experience that formulates the belief in the possibility of realisation of this sequence of events.
This paradox originates from a real phenomenon accessible to human senses and is not a thought experiment. Although the phenomenon is physically real, *the paradox is of theoretical nature*, since it could be clearly defined by no experiment and formulated only through experience as a consequence of long-term observation of nature.
This is a case of *mixed paradigms paradox* that is caused by the use of conceptually different modes of thought.

---
[1] http://en.wikipedia.org/wiki/List_of_paradoxes?

*The solution of Boltzmann's paradox*

Ludwig Boltzmann, one of the founders of statistical mechanics, did not believe that irreversible microscopic movement had a special characteristic that made it irreversible, other than being irreversible in its own right. The same goes for macroscopic movement that demonstrates its irreversibility by the fact that the reversible process is impossible (water spilt from a basin can "hardly" return to the basin), which, according to Boltzmann, is simply the natural order of things. Boltzmann's standpoint was that there simply was no solution to this paradox.

Macrosystems are entropically organised, and tend to maximal disorder. The nature of macrosystems is irreversible.

What Boltzmann managed to prove is that entropy of reversible processes amounts to the law of probability (clearly reflected in Planck's formula for entropy $S=k\cdot lnW$). Also, he demonstrated that entropy of reversible processes can amount to the low of statistical mechanics, but nowhere did he find the connection that produces irreversible processes from the reversible.

Lord Kelvin differentiates between the "abstract"[2] and the "physical"[3] dynamics, the former being reversible and the latter irreversible.[4] According to Lord Kelvin there is no real difference between these two mechanics and it only appears when statistical mechanics is introduced in mechanics.

There is no concrete solution of this paradox, apart from the cited opinions coming from authorities in physics which are speculative and philosophical rather than physical explanations.

## 2.Loschmidt's reversibility paradox

*A vessel with perfectly reflecting walls contains gas in non-equilibrium state (having a non-Maxwellian speed distribution). The state of the system can be described by Boltzmann's H-function. In time the gas will reach Maxwellian distribution. System state change can be described by $H_1$, $H_2$... $H_H$ progression, which is decreasing according to Boltzmann's H-theorem[5]. After the system equilibrium is reached all particle speed directions change. At first the new state corresponds to the second Boltzmann function $H_H$', which is as probable as $H_H$. Reversing the molecule speed direction should cause the gas to go through states of increasing character: $H_H$, $H_{H-1}$, ... $H_2$, $H_1$, which is in contrast to Botlzmann's H-theorem stating that H-function is decreasing. This refutes the Second principle of thermodynamics and shakes the foundations of thermodynamics.*

Joseph Loschmidt (1821–1895) published an article in 1876 in which he tried to prove that system equilibrium is possible without equalizing the temperatures of the elements that make up the system. This paradox, like the *Zermelo's paradox*, is based on the criticism of *Boltzmann's H-theorem*, and is one of a number of discussions that lead to the definition of statistical nature of the *Second principle of thermodynamics*. Like Lord Kelvin two years earlier, Loschmidt proposed an abstract situation of the reversal of system molecules speed direction.

---

[2] rational mechanics
[3] real physical processes
[4] Published in 1874 in his work *Kinetic theory of energy dissipation*.
[5] If function f(x, t) is the solution to the corresponding differential equation, and H is the solution to a certain integral of the function, than $dH/dt \leq 0$.

*The Loschmidt paradox* is based on the principle of *Newton's mechanics* equations stating that they do not change with the time direction change (time change initiates reversibility), and according to the classical thermodynamics principle all real processes are irreversible.

There is obvious contradiction between the law of entropy increase and the principle of Newton's classical mechanics, since it does not distinguish between past and future. This is the so-called *paradox of reversibility* that was formulated as Loschmidt's objection to Boltzmann's theory.[6]

This paradox does not originate from a real phenomenon, accessible to human senses. It is speculative and originates from a phenomenon formulated by thought. *This paradox is a thought experiment*. Also, it could be said *that the nature of the paradox is theoretical* since the paradox is based on the conflict between formulated theoretical concepts.

Considering that the paradox came up during the process of formulating the statistical model of thermodynamical phenomenon it can be classified as a transitional paradox.

*The solution of Loschmidt's paradox of reversibility*

The solution was found even before the paradox was defined, by Lord Kelvin in his 1874 work *Kinetic theory of energy dissipation*. Kelvin supposed that it was possible from temperature system equilibrium, by reversing system molecules direction, to reach the initial state of the system with uneven distribution of system temperature. His conclusion was that if there were a greater number of molecules in a system the time to reach the initial state is shorter. If the system's number of molecules tends to infinity, reaching non-equilibrium is considered impossible.

In 1877, Boltzmann replied to Loschmidt that gas molecules after H' state, if enough time had passed, would cross to the H state, and reach equilibrium by decreasing H-function, since the number of possible state distributions is far greater in equilibrium than non-equilibrium.

### 3. Zermelo's recurrent paradox

> *Poincare's recurrency theorem[7] suggests that Boltzmann H-function is periodic. According to Boltzmann's H-theorem H-function is decreasing[8] and it doesn't require the possibility of its returning to the initial state.*

Towards the end of the 19th century physicists employed the knowledge based on: Newton's classical mechanics, classical thermodynamics and *Maxwell's electrodynamics* (only just formed). Combined, Newton's classical mechanics and classical thermodynamics lead to *Loschmidt's and Zermelo's* paradox.

Poincare used *Zermelo's paradox* and proved that the initial states of Newton's equations would repeat suggesting, however, that this is not real irreversible process. The *recurrent paradox* is based on *Poincare's theorem*.[9]

---

[6] Translator's foreword to *Lectures on Gas Theory* by Ludwig Boltzmann.

[7] If a system is closed and has a limited number of states, ζ after a sufficiently long time the system will return to the state similar to the initial one. Time duration is measured by the so-called Poincare's number $10^\zeta$ (the number is so great that that the unit of time is irrelevant).

[8] $dH/dt \leq 0$

[9] *In an isolated limited system, with finite energy, confined to a finite volume, after a sufficiently long time, the system tends to return to a state similar to the initial one.*

Ernst Zermelo (Ernst Friedrich Ferdinand Zermelo, 1871-1953), a student of Max Planck and successor of his thermodynamic understanding of the world, published in 1896 an article entitled *On a Theorem of Dynamics and the Mechanical Theory of Heat* (Zermelo, 1896), in which he applied *Poicare's theorem* to the *Second principle of thermodynamics*.

Zermelo's conclusion does not originate from a real phenomenon, accessible to human senses. It is speculative and originates from a phenomenon formulated by thought. *This paradox is a thought experiment. The paradox is of theoretical nature*; it arises from a debate on the theoretical concepts of statistical nature of the explanation.

*Zermelo's recurrent paradox* is of distinct theoretical character and was formed in the process of defining Boltzmann's formulation of statistical physics as a critical review of the principles and opinions adopted by him. The paradox enables a better comparison of the formulations of statistical physics and it can be categorized as a *transitional paradox*.

*The solution of Zermelo's recurrent paradox*

Zermelo based his paradox on the view that if a process is periodic it could not be simultaneously irreversible. Thus, by applying *Poincare's theorem*, the term entropy in the *Second principle of thermodynamics* becomes absurd.

Twice does Boltzmann substantiate his views and provides a kind of a solution to *Zermelo's paradox*. He is conscious of the metaphysical character of the principle that all was created from the least likely state and physical tendency to change the state of a system (Universe) and a subsystem to a more probable state. According to him, *Poincare's theorem* is "less general" than the adopted principle and although recurrency is to be expected after a long enough period of time, generally speaking, system entropy increases.

### 4. Mpemba effect

> *In two equal containers with equal volume of water at non-drastically different temperatures, the freezing of their contents begins using the same process. Under certain conditions, the water that was initially warmer will start to freeze first.*[10]

The phenomenon that warmer water freezes quicker than colder was know back in Aristotle's days (384-322 BC)[11]. It was discussed during the early middle ages by Roger Bacon (1214-1294)[12], then Giovanni Marliani[13] in the 15th century, and in the late middle ages it was described by Francis Bacon (1561-1626)[14] and Rene Descartes (1596-1650)[15]. In a strange way the phenomenon was neglected and forgotten by the scientific public until, thanks to a Tanzanian student Mpemba (Erasto B. Mpemba) it returned to public attention in 1969. This is why the phenomenon is also known as the *Mpemba effect*. The reason why warmer water freezes quicker than colder is not yet known. Since common sense tells us that it is illogical the effect is also called a paradox.

---

[10] People know from experience that a frozen windscreen is defrosted with cold water rather than hot, since hot water would freeze quicker, before the windscreen is wiped clean.
[11] In chapter 12 of the first book of *Meteorologica*.
[12] *Opus Majus* (Greater work).
[13] M. Clagett, *Giovanni Marliani and the late medieval physics*, AMS Press Inc., New York, 1967, p.79.
[14] *Novum Organum*.
[15] *Discourse on Method, Optics, Geometry, and Meteorology*

This renewed interest in why warmer water freezes quicker than colder has its own anecdote now that Mpemba rediscovered it. Namely, while Mpemba was still in secondary school student of Eugene Marschal in Mkwawa School in Iringa[16] in 1964, he was making ice cream. He mixed hot milk with sugar and rather than wait for it to cool placed it in the refrigerator. To his surprise he noticed that his hot milk ice cream had frozen quicker than the ice cream of other students whose milk wasn't hot. When he asked his science teacher to explain it, he couldn't. The first person to take him seriously was Denis G. Osborne in 1969, who repeated the experiment after Mpemba inquired about the reason why warmer water froze quicker than colder. They described this effect together and returned it to the focus of scientific public's attention. (Mpemba & Osborne, 1969).

As it happens, the same year Kell (Kell. G. S) published his work, independently from Mpemba and Osborne, in which he endeavoured to explain the phenomenon by evaporation. He was not familiar with Denis Osborne's experiment which showed that evaporated mass is not enough to explain the effect.

This paradox originates from a real phenomenon, accessible to human senses, and is not a *thought experiment*. The above mentioned experiment entails that this is an *experimental paradox*.

Since this is a paradox without an explanation of the temperature change speed in different physical states of water, clearly hierarchically differentiated, it could be classified as a *hierarchical paradox*.

*The solution of Mpemba effect*

In order to solve Mpemba effect it is necessary to define precisely the initial conditions (water mass, shape and type of the vessel, temperature difference, heat convection in the water, volume of air in the water, freezing method, surrounding system,...) in which the effect is expected to be demonstrated. In the case of drastic difference in water temperature or volume of water the effect will not occur. It needs to be precisely defined whether the time interval observed will be until the beginning of freezing or until the complete freezing of the entire volume of water.

The problem is that *Mpemba effect* is inconsistent with modern heat theory.

One of the explanations is that since at first we have equal volumes of water at different temperatures, the volume of water at the higher temperature evaporates more, so that at the moment of freezing the volumes of water will not be equal and the smaller amount of water will freeze quicker. Certain authors claim that evaporation process alone is insufficient to explain *Mpemba effect*.

It is generally believed that there is an infinite number of ways to combine relevant experiment parameters that set up the conditions under which Mpemba effect will or will not apply. One cannot say that *Mpemba effect* has been solved, since the values of the parameters within which warmer water will freeze quicker than cold have not been precisely determined.

---

[16] Tanzania

## 5. Maxwell's demon paradox

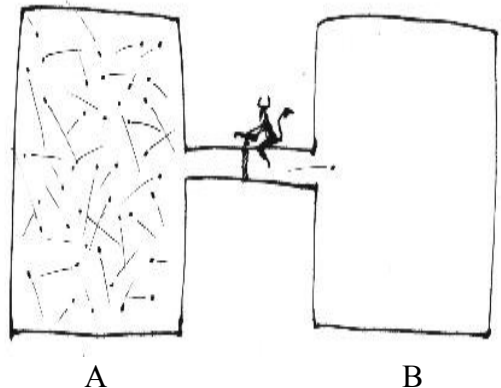

A                                      B

> *"Imagine a creature capable of following every single molecule along its path. Such creature, whose characteristics would basically be as final as our own, would be capable of something that we are not. This is because the molecules in an air-filled container, at a regular temperature, move at velocities that are not regular at all, although the average velocity of a large, randomly selected, number of molecules is almost completely regular. Assume that such a container would be divided into two compartments, A and B, by a screen with a small hole in it. The creature that can see individual molecules would open and close the hole so that only the faster molecules may pass from A to B and only the slower from B to A. Thus, the temperature in B would increase and the temperature in A would diminish without work, which contradicts the Second law of thermodynamics"*[17]

*Maxwell's demon paradox* is given here in its original form. James Maxwell published the paradox in 1871 in a footnote in a textbook called *Theory of Heat*, and the paradox was first mentioned in December of 1867 in Maxwell's letter to Tait (Peter Guthrie Tait, 1831-1901). The "demon" will be introduced later[18] instead of the "being" that is arranging molecules according to their speed, that "*considers the way in which, when two objects are in contact, the wormer object takes over heat from the colder without external intervention.*"[19]

This paradox does not originate from a real phenomenon, accessible to human senses. It is speculative and originates from a phenomenon formulated by thought. *This paradox is a thought experiment*. (Cucić, 2001) *The paradox is of theoretical nature* created as a thought speculation in order to help in presenting a theoretical concept.

What makes this fictitious experiment paradoxical is the approach to the *Second principle of thermodynamics* that heat can never be transferred from a colder to a warmer object without external intervention. Maxwell found a way to "shake the foundations" of classical thermodynamics owing to different principle of addressing the problem of heat transfer.

*This paradox appeared as a consequence of paradigm mixing*. In this case the paradigms of classical physics and quantum physics are mixed. A quantum mechanical problem is approached from the viewpoint of classical physics.

---

[17] (Mlađenović, 1989, p.207-208), adopted from J.C. Maxwell, Theory of Heat, Longmans; London, 1891, section "Limitations of the Second Law of Thermodynamics".
[18] by Joseph John Thompson (1856-1940)
[19] "From the Maxwell demon to the granular clock", p. 2-4

*The solution of Maxwell's demon paradox*

The thought experiment is paradoxical because the "being" that is arranging the molecules belongs to a world of classical physics. The "being" acts according to the rules of the macrocosm in the microcosm. If the paradox is viewed from a quantum mechanical standpoint the existence of the "demon" must be understood as active intervention, since it does work while choosing molecules.

### 6. Heat death paradox (Clausius paradox)

*Assuming that universe is eternal a question arises: How is it that thermodynamic equilibrium has not been long achieved?*

Heat death paradox, otherwise known as Clausius paradox and Thermodynamic paradox, is founded on the basic assumption that every system tends to achieve thermodynamic equilibrium. This paradox is based upon the classical model of the universe in which the universe is eternal. C*lausius' paradox* is *paradox of paradigm*. It was necessary to amend the fundamental ideas about the universe, which brought about the change of the paradigm. The paradox was solved when the paradigm was changed. The paradox which is valid in the classical stationary model of the universe is not so in Fridman's nonstationary relativistic model. This is a solved paradox and therefore is an *exparadox*.

The paradox was based upon the rigid mechanical point of view of the *Second principle of thermodynamics* postulated by Rudolf Clausius according to which heat can only be transferred from a warmer to a colder object. If the universe was eternal, as claimed in the classical stationary model of the universe, it should already be cold.

This paradox originates from a real phenomenon accessible to human senses and is not a thought experiment but a real sensory observation. There is no experiment that could demonstrate this paradox. *The paradox is of theoretical nature*.

*The solution of the Heat Death paradox*

According to recent cosmological theories the universe is not eternal and it began some 15 billion years ago. In view of this, the solution of the *Heat death* p*aradox* is that thermodynamic equilibrium has not been achieved because not enough time has passed.

### 7. Gibbs' paradox (paradox of mixing)

*Assume that there is a vessel, divided by a barrier into two equal compartments containing the same gas, in equal quantities, at equal temperature and pressure. The removal of the barrier and mixing of gas from each compartment causes increase in system entropy.*

This paradox is based in the mixing of gases, providing that practically nothing has changed (gas quantity, temperature, pressure, volume). When the barrier is removed each of the two quantities of gas spreads over the entire volume of the vessel changing the system entropy, which is easy to calculate. Under equal conditions the entropy of separate compartments is equal. Prior to the removal of the barrier the system entropy is 2S, after the barrier has been removed it is higher than 2S. This phenomenon is called *entropy of mixing*. Paradox of mixing comprises of calculating the entropy of two thermodynamic systems, before and after their contents are mixed. The paradox was formulated by Willard Josiah Gibbs in 1875. Gibbs first developed the paradox of mixing in 1861. The paradox of mixing is the former name of Gibbs' paradox. The paradox is a "precognitive" introduction into Bose and Fermi statistics.

This paradox originates from a real phenomenon accessible to human senses and is not a thought experiment. It is a practicable laboratory experiment for paradox interpretation, although its essence lies in its explanation.

The paradox was formulated and solved by Gibbs himself. He pointed out the possibility of "misunderstanding" and at the same time explained why it might occur. Thus, this is an *exparadox*.

Considering that *Gibbs' paradox* exists when the phenomenon is viewed from the standpoint of classical statistic mechanics, and that the paradox does not exist if it is explained by quantum statistic, this paradox can be classified as a *paradox of paradigm*. Changing the basic paradigmatic principles brings about the solution of the paradox.

*The solution of Gibbs' paradox*

It is the fundamental assumption in classical statistic mechanic that identical particles are distinguishable from one another. System states, resulting from permutation of identical particles, are indistinguishable from one another. From the point of view of classical statistic entropy has no additive characteristics.[20] Gibbs concluded that classical statistic mechanics should be treated as a border case, and that the fundamental assumption should be that identical particles cannot be told apart[21]. In that case the paradox of mixing does not exist since no work is done by the movement of particles in the act of invisible mixing.

**Conclusion**

This work presented seven paradoxes, two of which were paradoxes of classical thermodynamics (Mpemba effect and Clausius' paradox), and five were created as a consequence of a statistic approach to thermodynamic phenomena (Boltzmann's paradox, Loschmidt's paradox of reversibility, Zermelo's recurrent paradox, Maxwell's demon paradox and Gibbs' paradox). All the paradoxes date from the 19$^{th}$ century except Mpemba effect which was known in ancient Greece.

The paradoxes of thermodynamics and statistical physics are most frequently formed, as many as four out of seven, by faulty paradigm and its change brought about their solution (Boltzmann's paradox, Clausius' paradox, Maxwell's demon paradox and Gibbs' paradox). Two of the seven paradoxes are transitional – created in the process of formulating a theory based on an unchanged paradigm (Loschmidt's paradox of reversibility, Zermelo's recurrent paradox). One paradox is hierarchical – created by changes in system states (Mpemba effect).

---

[20] Total system entropy in quantum statistic is calculated by simple addition of the entropies of individual vessels: $(N_A+N_B) \cdot k \cdot \ln(2)$, where $N_A$ and $N_B$ are a total number of particles in vessels A and B.
[21] Which is the basic principle of quantum statistics.

Five of the seven paradoxes could be said to have "solid" solutions and that they meet the criteria necessary to call a paradox solved (Clausius' paradox, Loschmidt's paradox of reversibility, Zermelo's recurrent paradox, Maxwell's demon paradox and Gibbs' paradox). For two paradoxes there are explanations that suggest that their solutions exist but that they are still not convincing enough to be completely accepted (Mpemba effect and Boltzmann's paradox).

Three paradoxes are thought experiments that have no foundation in real experiments (Loschmidt's reversibility paradox, Zermelo' recurrent paradox and Maxwell's demon paradox), while four of the seven paradoxes are founded in real physical observations (Mpemba effect, Clausius' paradox, Boltzmann's paradox and Gibbs' paradox). Two paradoxes directly follow from real experiments (Mpemba effect and Gibbs' paradox) while as many as five originate in theoretical deliberation (Boltzmann's paradox, Loschmidt's reversibility paradox, Zermelo' recurrent paradox, Maxwell's demon paradox and Clausius' paradox). This analysis has been presented in tabular and graphical form.

The paradoxes in thermodynamics and statistical physics are different and posses very entangled characteristics, which makes it impossible to single out a special property that would set them apart from the paradoxes of other branches of physics.

## Acknowledgements

I wish to express my gratitude to academic sculptor Zoran Maleš for the drawing *Maxwell's demon*.

| Pseudo paradox | Paradox of idealization | Hierarchical paradox | Transitions paradox | Paradox of assumption | Paradox of paradigm |
|---|---|---|---|---|---|
| | | Mpemba effect | Loschmidt's reversibility paradox | | Boltzmann's paradox |
| | | | Zermelo's recurrent paradox | | Maxwell's demon paradox |
| | | | | | Clausius paradox |
| | | | | | Gibbs' paradox |

*Table 1*. Table representation of paradoxes in Thermodynamics and Statistical Physics.

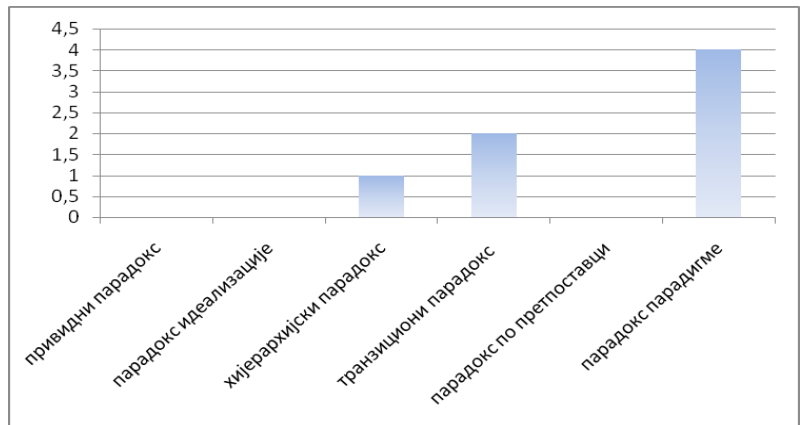

Graphic 1. Graphical representation of paradoxes in Thermodynamics and Statistical Physics.

| *Exparadox* | Loschmidt's reversibility paradox | Zermelo's recurrent paradox | Maxwell's demon paradox | Clausius paradox | Gibbs' paradox |
|---|---|---|---|---|---|
| **Unsolved paradox** | Boltzmann's paradox | Mpemba effect | | | |

*Table 2.* Table representation of solvability of paradoxes in Thermodynamics and Statistical Physics.

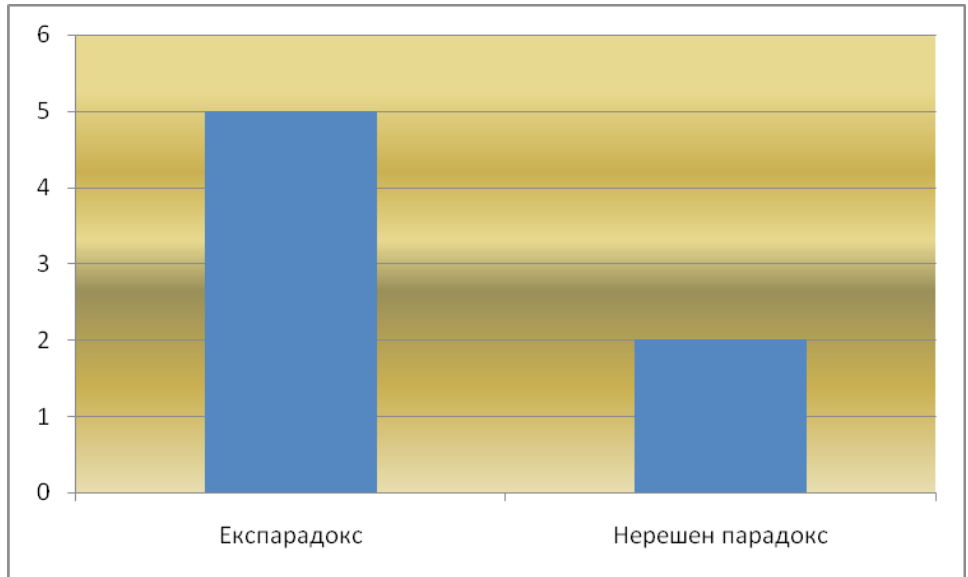

Graphic 2. Graphical representation of solvability of paradoxes in Thermodynamics and Statistical Physics.

| **Thought experiment** | Loschmidt's reversibility paradox | Zermelo's recurrent paradox | Maxwell's demon paradox | |
|---|---|---|---|---|
| **Real sensory observation** | Boltzmann's paradox | Gibbs' paradox | Clausius paradox | Mpemba effect |

*Table 3.* Table representation of thought paradoxes in Thermodynamics and Statistical Physics.

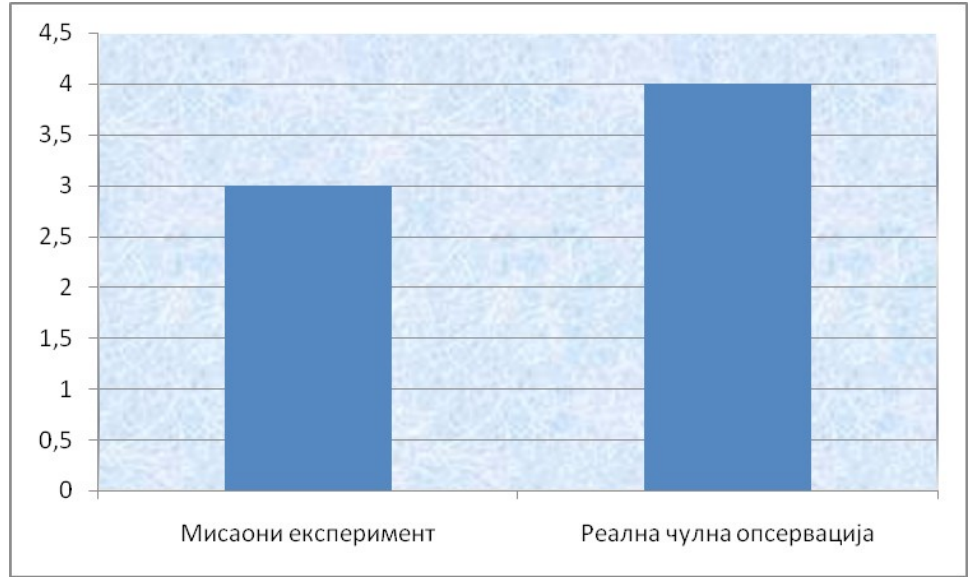

Graphic 3. Graphical representation of thought paradoxes in Thermodynamics and Statistical Physics.

| **Експериментални парадокс** | Gibbs' paradox | Mpemba effect | | | |
|---|---|---|---|---|---|
| **Теоријски парадокс** | Boltzmann's paradox | Zermelo's recurrent paradox | Maxwell's demon paradox | Clausius paradox | Loschmidt's reversibility paradox |

*Table 4.* Table representation of experimental and theoretical paradoxes in Thermodynamics and Statistical Physics..

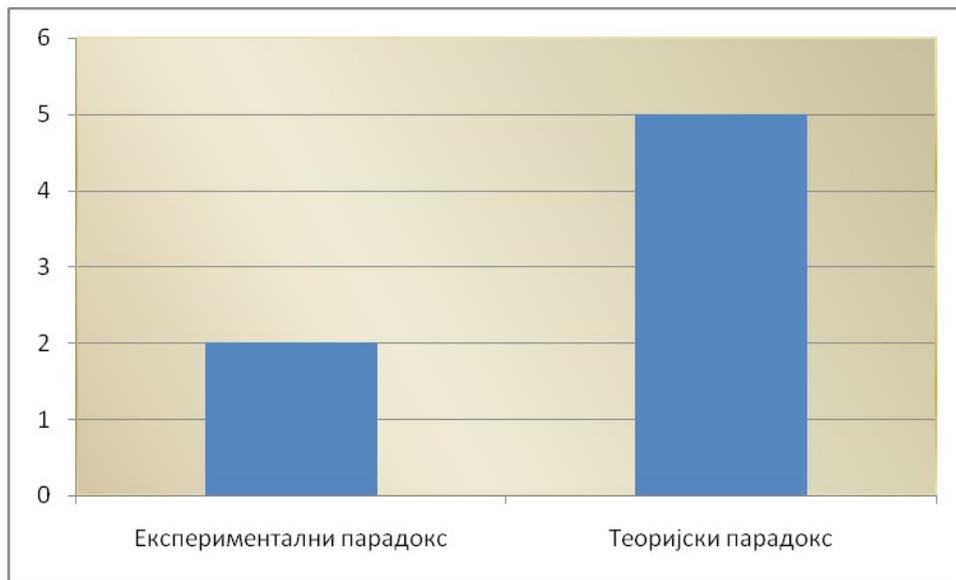

*Graphic 4.* Graphical representation of experimental and theoretical paradoxes in Thermodynamics and Statistical Physics.

|  | **Boltzmann's paradox** | **Loschmidt's reversibility paradox** | **Zermelo's recurrent paradox** | **Mpemba effect** | **Maxwell's demon paradox** | **Clausius paradox** | **Gibbs' paradox** |
|---|---|---|---|---|---|---|---|
| **type of paradoxes** | Paradox of paradigm | Transitions paradox | Transitions paradox | Hierarchical paradox | Paradox of paradigm | Paradox of paradigm | Paradox of paradigm |
| **solvability of paradoxes** | unsolved | exparadox | exparadox | unsolved | exparadox | exparadox | exparadox |
| **thougth of paradoxes** | theory | theory | theory | experimental | theory | theory | experimental |
| **experimental of paradoxes** | opservation | thought | thought | opservation | thought | opservation | opservation |

*Table 5.* Tabelar view of Paradoxes in Thermodynamics and Statistical Physics.